# A missing high-spin molecule in the family of cyano-bridged heptanuclear heterometal complexes, $[(LCu^{II})_6Fe^{III}(CN)_6]^{3+}$, and its $Co^{III}$ and $Cr^{III}$ analogues, accompanied in the crystal by a novel octameric water cluster. Magnetic and NMR studies

Catalin Maxim,[a] Lorenzo Sorace,[b] Panchanana Khuntia,[c,d] Augustin M. Madalan,[a] Alessandro Lascialfari,[c] Andrea Caneschi,[b] Yves Journaux[e], Marius Andruh[a]

[a] *Inorganic Chemistry Laboratory, Faculty of Chemistry, University of Bucharest, Str. Dumbrava Rosie nr. 23, 020464-Bucharest, Romania*
[b] *Laboratory of Molecular Magnetism, Dipartimento di Chimica e UdR INSTM di Firenze, Università degli Studi di Firenze, Polo Scientifico, Via della Lastruccia 3, 50019 Sesto Fiorentino, Firenze, Italy*
[c] *Dipartimento di Fisica"A.Volta" e Unita' CNISM-CNR, Universita' di Pavia, Pavia, Italy*
[d] *Department of Physics, Indian Institute of Technology Bombay, Powai, Mumbai-400076, India*
[e] *Institut Parisien de Chimie Moléculaire, UPMC and CNRS UMR7201, case courrier 42, UPMC-Paris 6*

Three isostructural cyano-bridged heptanuclear complexes, $[\{Cu^{II}(saldmen)(H_2O)\}_6\{M^{III}(CN)_6\}](ClO_4)_3 \cdot 8H_2O$ (M = $Fe^{III}$ **2**; $Co^{III}$, **3**; $Cr^{III}$ **4**), have been obtained by reacting the binuclear copper(II) complex, $[Cu_2(saldmen)_2(\mu\text{-}H_2O)(H_2O)_2](ClO_4)_2 \cdot 2H_2O$ **1**, with $K_3[Co(CN)_6]$, $K_4[Fe(CN)_6]$, and, respectively, $K_3[Cr(CN)_6]$ (Hsaldmen is the Schiff base resulted from the condensation of salicylaldehyde with N,N-dimethylethylenediamine). A unique octameric water cluster, with bicyclo[2,2,2]octane-like structure, is sandwiched between the heptanuclear cations in **2**, **3** and **4**. The cryomagnetic investigations of compounds **2** and **4** reveal ferromagnetic couplings of the central $Fe^{III}$ or $Cr^{III}$ ions with the $Cu^{II}$ ions ($J_{CuFe}$ = +0.87 cm$^{-1}$, $J_{CuCr}$ = +30.4 cm$^{-1}$). The intramolecular Cu···Cu exchange interaction in **3**, across the diamagnetic cobalt(III) ion, is -0.3 cm$^{-1}$. The solid-state $^1$H-NMR spectra of compounds **2** and **3** have been investigated.

**Introduction**

Perhaps the richest chemistry in the realm of heterometal complexes is the one based on self-assembly processes involving polycyanometallate tectons and various assembling complex cations. Indeed, a huge number of heterometallic complexes have been synthesized over the last 15 years.[1] The main interest in such compounds arises from their exciting magnetic properties. In this respect, the most employed building-blocks are paramagnetic cyano-complexes: [M(CN)$_6$]$^{3-}$ (M = Fe$^{III}$, Cr$^{III}$); [M(CN)$_8$]$^{3-}$ (M = Mo$^V$, W$^V$); [Mo(CN)$_7$]$^{4-}$. More recently the family of cyano-bridged heterometallic complexes has been enriched by using heteroleptic anionic species, such as [M(AA)(CN)$_4$]$^-$ (M = Fe$^{III}$, Cr$^{III}$, AA = 2,2'-bipyridine, 1,10-phenanthroline, 2-aminomethyl-pyridine).[2] The assembling cations are either fully hydrated metal ions or complex species in which the metal ion carry a blocking ligand. In the first case, Prussian-Blue phases are obtained. In the second one, depending on the denticity of the blocking ligand, a large variety of bimetallic complexes, ranging from high-nuclearity clusters to coordination polymers with various dimensionalities and network topologies, are obtained. The number of organic molecules which can be used to block several coordination sites at the assembling cations is unlimited. They can have various denticities, ranging from monodentate to pentadentate, and can be linear, tripodal or macrocyclic.[3]

The bimetallic oligonuclear complexes are very useful because they are good models for studying the sign and magnitude of the exchange interaction mediated by the cyano bridge. Moreover, those complexes with a high spin ground state and a large uniaxial magnetic anisotropy show very peculiar magnetic properties like slow relaxation of the magnetization and quantum tunnelling of the magnetization.[4] Taking the case of [M(CN)$_6$]$^{3-}$ tectons, the highest nuclearity that can be reached involving the six cyano groups, as bridges, is seven: [M{CN)M'L}$_6$]. The second metal ion, M', carries a polydentate blocking ligand, L. In a series of papers, Marvaud, Verdaguer et al. have shown that, by changing the stoichimetry in the reaction between the hexacyanometallate and the cationic complex, the nature of the blocking ligand, or the counterion, it is possible to obtain complexes with nuclearities ranging from two to seven.[3a,b]

In spite of the richness of the chemistry of cyano-bridged heterobimetallics, the number of heptanuclear clusters of the type [M{CN}M'L}$_6$] is limited to only few examples. Two of the very first heptanuclear clusters, [Cr$^{III}$Mn$^{II}_6$] and [Cr$^{III}$Ni$^{II}_6$], have been obtained by Mallah and Verdagurer, using [Cr(CN)$_6$]$^{3-}$ as a template.[5] Several other heptanuclear complexes are also obtained from hexacyanochromate(III).[3a,6] Surprisingly, the search of the literature shows that no heptanuclear complexes derived from [Fe(CN)$_6$]$^{3-}$ are described. All the reported attempts to obtain [Cu$^{II}_6$Fe$^{III}$] cyano-bridged clusters failed because of reduction of Fe$^{III}$ to Fe$^{II}$.[7] Here we report the synthesis, crystal structure and magnetic properties of the first [Cu$^{II}_6$Fe$^{III}$] cyano-bridged complex, as well as the isotructural cobalt(III) and chromium(III) derivatives.

## Results and Discussion

The Cu$^{II}$ – Fe$^{III}$ (low spin) pair is particularly interesting, because the orthogonality of the magnetic orbitals should lead to a ferromagnetic coupling. Frequently, copper(II) exhibits a square-pyramidal stereochemistry, the magnetic orbital, d$_{x2-y2}$, being localized in the basal plane. Consequently, in order to be efficient, the magnetic interaction with the other metal ion must occur along a bridge connecting the basal plane of the coordination polyhedron of copper(II) with the other metal ion. In this respect, we have chosen as a blocking ligand a tridentate Schiff-base, obtained from the condensation reaction of salicylaldehyde with N,N-dimethyl-ethylenediamine (Hsaldmen), whose reaction with copper(II) perchlorate affords a binuclear complex, [Cu$_2$(saldmen)$_2$(μ-H$_2$O)(H$_2$O)$_2$](ClO$_4$)$_2$·2H$_2$O **1**. Compound **1**, with the copper ions weakly bridged by the aqua ligand, can be employed as a precursor in obtaining heterometallic complexes. Indeed, the reaction between **1** and K$_4$[Fe(CN)$_6$] yields the heptanuclear complex [{Cu(saldmen)(H$_2$O)}$_6${Fe(CN)$_6$}](ClO$_4$)$_3$·8H$_2$O **2**. During the reaction, Fe$^{II}$ got oxidized to Fe$^{III}$. The infrared spectrum of compound **2** shows one band at 2104 cm$^{-1}$, which is characteristic for the cyano group connecting iron(III) and another metal ion.[1a] We recall that for complexes containing the hexacyanoferrate(II) ion, this band lies below 2100

cm$^{-1}$.[1a] The isostructural Co$^{III}$ and Cr$^{III}$ complexes, [{Cu(saldmen)(H$_2$O)}$_6${Co(CN)$_6$}](ClO$_4$)$_3$·8H$_2$O **3**, and [{Cu(saldmen)(H$_2$O)}$_6${Cr(CN)$_6$}](ClO$_4$)$_3$·8H$_2$O **4**, have been obtained in a similar way, replacing K$_4$[Fe(CN)$_6$] with K$_3$[Co(CN)$_6$] and, respectively K$_3$[Cr(CN)$_6$]. The characteristic $\nu_{CN}$ bands in compounds **3** and **4** are located at 2184 cm$^{-1}$ and, respectively, 2183 cm$^{-1}$.

**Crystal structures**

The crystal structures of compounds **1**, **2**, **3**, and **4** have been solved. Selected bond distances and angles for the four compounds are collected in Table 1. There are two important features revealed by the crystallographic analysis of **1**: the copper ions are bridged by an aqua ligand, and each copper ion is coordinated into the fourth basal position by a terminal aqua ligand, which, in principle, can be substituted by a bridging group (Fig. 1). This compound is very similar to another aqua-bridged binuclear copper(II) complex reported recently by Mitra, Sutter et al., which has been obtained from a slightly different Schiff-base, derived from hydroxyacetophenone, instead of salicylaldehyde.[8] The copper ions are pentacoordinated, with a square pyramidal stereochemistry. The basal plane is described by the tridentate anionic NNO Schiff base and one aqua ligand, with copper – donor atoms distances varying between 1.904(3) and 2.037(4) Å. The apical positions of the two copper ions are connected by the aqua bridge [Cu1 – O2w = 2.477(3) Å], with a distance between the copper ions of 4.060 Å. Intramolecular hydrogen bonds are established between the terminal H$_2$O ligand from one moiety and the phenoxo oxygen from another one (O1···O1w = 2.664 Å).

The crystal structure of compound **2** (trigonal *R*-3*c* space group) consists of heptanuclear cations, [{Cu(saldmen)(H$_2$O)}$_6${Fe(CN)$_6$}]$^{3+}$, uncoordinated perchlorate ions and crystallization water molecules. The structure of the heptanuclear cation is illustrated in Fig. 2. The six copper(II) ions are connected to the cyano groups [identical Cu – N distances, 1.932(5) Å, and Cu – N - C angles of 171.7(5)º]. The copper ions display a square-pyramidal geometry, with the aqua ligands coordinated into the apical position [2.395(6) Å]. The basal plane is described by the tridentate Schiff base anion and one nitrogen atom arising from the cyano bridge.

The analysis of the packing diagram for crystal **2** reveals the organization of octameric water clusters with a unique supramolecular architecture. The systematic investigation of water clusters of various sizes is crucial for gaining insights into the structure and properties of bulk water,[9] for the understanding of the role played by the small water clusters in stabilizing and functioning of biomolecules,[10] or of the key role played by the water clusters in designing novel metal-organic materials.[11] Water clusters of various nuclearities, $(H_2O)_n$, have been characterized so far.[12] Spectacular large water clusters with a spherical architecture (nano-drops) were characterized by Müller et al.[13] Since the water molecules can act as supramolecular tetrahedral synthons, their connection through hydrogen bonds can lead to supramolecular architectures that have a correspondent in hydrocarbon chemistry. For example, water clusters with cyclobutane,[12g] cyclopentane,[12r] or cyclohexane-like conformations[12i,t] are well known. In our case, the eight water molecules form a unique supamolecular cluster with a bicyclo[2,2,2]octane-like structure (Fig. 3). It is also interesting to notice that this cluster, being predicted by theoretical calculations several years ago,[9c] represents a fragment of the $I_h$ ice. The geometrical parameters associated to the hydrogen bonds are given in Table 2. The topologies of the cyclic octameric water clusters can follow the conformations of the cyclooctane (crown, boat, boat-chair), which are also constructed from tetrahedral synthons. Another possible topology is the cage-like one, with the water molecules disposed at the corners of the cube. The very first crystallographically characterized octameric water cluster exhibits this last topology.[14] Another $(H_2O)_8$ cluster was observed by Atwood *et al.*, and shows a cyclooctane-like boat conformation.[15] More recently, some of us have described a crwon-like octameric cluster, with a $D_{4d}$ symmetry.[16] Compounds **3** and **4** are isostructural with compound **2**.

## Magnetic properties

We begin our discussion of the magnetic properties of the three heterometallic complexes with the temperature dependence of $\chi T$ for compound **3**, which is illustrated in Fig. 4a. ($\chi$ represents the molar magnetic susceptibility). The highest temperature value (2.42 cm$^3$mol$^{-1}$K at 160 K) corresponds to that expected for six uncoupled copper(II) ions with g = 2.07. The $\chi T$ product remains constant down to 25 K, then it decreases abruptly, indicating that antiferromagnetic interactions are active. By fitting these data with a Curie-Weiss law (inset of Figure 4a), we obtained $C = 2.44$ cm$^3$mol$^{-1}$K and $\theta = -0.56$ K. The low value of the Weiss term, $\theta$, is in agreement with weak antiferromagnetic interactions between the copper(II) ions. These interactions can in principle be either intramolecular (across the diamagnetic cobalt(III) ion) or intermolecular (mediated by hydrogen bonds involving the aqua ligand). These last interactions should however be much weaker than the intramolecular ones, since the magnetic orbital of the copper(II) ion, $d_{x2-y2}$, is not oriented towards the apical aqua ligand. Consequently, we will consider only the intramolecular coupling of the copper ions along the Cu-Co-Cu paths. The average value of the exchange parameter, $J_{CuCu}$, has been obtained by using the simple Mean Field Approximation for the Weiss temperature, $\theta = zs(s + 1)J_{CuCu}/3k_B$, that gives $J_{CuCu} = -0.3$ cm$^{-1}$ ($z = 5$, $s = ½$, $k_B$ = Boltzmann constant). This value is close to those found in other cyano-bridged Cu$^{II}$ – Co$^{III}$ complexes,[17] but it has to be kept in mind that it is an average value, since the two groups of exchange coupling paths feature a Cu-Co-Cu angle of 180° and 90° respectively. However a fit using complete diagonalization of the spin Hamiltonian, with two different coupling constants,[18] evidenced an almost complete correlation between their values, thus leading to unreliable results. For this reason, we keep the average value obtained by MFA as a reasonable guess of the coupling constant value.

The $\chi T$ versus $T$ curve of **2** is represented in Fig. 4b. At 250 K, the value of the $\chi T$ product is 2.71 cm$^3$mol$^{-1}$K, which corresponds to seven non-interacting $S = ½$ metal ions with $g > 2$. By lowering the temperature, $\chi T$ remains constant down to 20 K, then it weakly increase to reach 2.81cm$^3$mol$^{-1}$K at 2 K. This is in agreement with expectations of a ferromagnetic coupling, since the unpaired electron density for pseudo-octahedral low-spin iron(III) ion is located in $d_{xy}$, $d_{yz}$, $d_{zx}$ orbitals, that are quasi-degenerate and $\pi$ in character, while the unpaired electron of the square pyramidal copper(II) ion is located in the $d_{x2-y2}$ orbital, which is $\sigma$ in character. The magnetic orbitals on the interacting centers are then orthogonal each other and a ferromagnetic interaction is then expected. It is worth noting that these data point to a negligible orbital contribution for the low spin Fe$^{III}$ since the room temperature value is close to the spin-only one and, further, no appreciable temperature dependence of the $\chi T$ product is observed in the high temperature range. A comparable behaviour has been recently reported for a linear, cyanide bridged, Cu-Fe-Cu complex, for which the almost complete quench of the angular momentum was attributed to the peculiar geometrical distortion of $Fe(CN)_6$ unit.[17]

The fit to the data for compound **2** was then performed by taking into account the interactions between $Cu^{II}$ and $Fe^{III}$, as well as the antiferromagnetic interactions occurring between the copper(II) ions, which were fixed to the average value obtained for the cobalt(III) derivative, **3,** and neglecting the orbital contribution of the low-spin Fe(III)**.** The best fit curve was obtained by using the following values: $J_{CuFe} = +0.87$ cm$^{-1}$; $g_{av} = 2.04$ ($J_{CuCu}$ fixed at -0.3 cm$^{-1}$). As mentioned above the ferromagnetic character of $Fe^{III}$-$Cu^{II}$ interaction is in agreement with expectations. The $Fe^{III}$-CN-$Cu^{II}$ exchange interaction was found ferromagnetic with several other compounds, with either discrete or extended structures.[19]

The temperature dependence of the $\chi T$ product for compound **4** is illustrated in Fig. 4c. The magnetic interaction between $Cr^{III}$ and $Cu^{II}$ is expected to be ferromagnetic, because the magnetic orbitals of these ions are orthogonal. Indeed, below 160 K, $\chi T$ increases continuously reaching a value of 13.17 cm$^3$mol$^{-1}$K at 13 K, then it decreases (the room temperature value of $\chi T$ is 6.63 cm$^3$mol$^{-1}$K, and corresponds to one $Cr^{III}$ and six $Cu^{II}$ uncoupled ions). Followiong the same fitting procedure as for compound **2**, we obtained $J_{CuCr} = +30.4$ cm$^{-1}$. Similar ferromagnetic couplings between $Cu^{II}$ and $Cr^{III}$ ions were found in other cyano-bridged complexes.[3a, 19a,d,k,m,p,v]

## NMR data in [Cu$_6$Fe] and [Cu$_6$Co] complexes

For a more complete set of $^1$H NMR data, the reader is referred to our previous wotk,[20] here we limit our discussion to spectra shapes and linewidths. The $^1$H NMR spectra have been found to broaden progressively by decreasing temperature, with an appreciable frequency shift of the peak (resulting in a double structure) from the Larmor frequency. The broadening of the $^1$H NMR spectra is mainly of dipolar origin and the shift of the line is due to a field induced paramagnetic shift. The spectra at low temperatures (1.6-20 K), which give information regarding the local hyperfine field at different proton sites due to magnetic moments of $Cu^{II}$ and $Fe^{III}/Co^{III}$, were simulated with two Gaussian functions and are shown in Fig. 5 and Fig. 6.

The temperature dependence of the proton NMR linewidth (full width at half maximum, FWHM) referred to the central line only, follows a simple Curie-Weiss behavior and is directly proportional to the magnetic susceptibility $\chi$ as shown in Fig. 7. The line shape and width is determined by three contributions: (i) nuclear dipole-dipole interaction; (ii) dipolar hyperfine interactions of the hydrogen nuclei with the neighboring magnetic ions; (iii) a direct contact term arising from interaction of the nuclei with the local magnetic moments of $Cu^{II}$ and $Fe^{III}/Co^{III}$, coming from the hybridization of the proton s-wave function with the d-wave function of magnetic ions. The last term is responsible of the line shift at low temperature. On the other hand, the inhomogeneous broadening is due to the presence of many non-equivalent protons and to the powder's distribution (different orientation of crystallites with respect to magnetic field). The narrowing of NMR line width at high temperature can be attributed to the self diffusion of protons, to the decrease of proton dipolar interaction due to interstitial diffusion and to the sensitivity of NMR to CN bridges molecular motion.[21,22]

To analyze the behaviour of the shifted line, we remind that the paramagnetic shift is defined as $K_{ps}=(\nu_R-\nu_L)/\nu_L$, where $\nu_R$ is the resonance frequency and $\nu_L$ is the unshifted Larmor frequency of isolated proton.[21] The temperature dependence of the paramagnetic shift follows a Curie-type behaviour with Curie constant $C = 2.71$ cm$^3$mol$^{-1}$K in [Cu$_6$Fe] and $C = 2.44$ cm$^3$mol$^{-1}$K in [Cu$_6$Co], consistent with those obtained from magnetic susceptibility data.[20] From $K_{ps}$ we have recently calculated a value of 502 G for the hyperfine magnetic field at the proton site for hydrogen bonded to the $Cu^{II}$ for [Cu$_6$Fe] and of 462.4 G in the case of [Cu$_6$Co]. Taking into account that the dominant hyperfine interaction is dipolar, the averaging over all proton sites and all particle orientations yields a broadening of the line width proportional to the external magnetic field and the magnetic susceptibility. This can be seen easily by the linear field dependence of the FWHM (calculated for the central line only) at $T = 300$ K, 77 K and 4.2 K for [Cu$_6$Fe] and at $T = 300$ K, 77 K for [Cu$_6$Co], see Figs.8 and 9. The solid lines in Figs. 8 and 8, represent a linear fit to :

FWHM $\approx \Delta\nu_d + \chi_{loc}H$ \hfill (1)

where $\chi_{loc}$ represents the "local" susceptibility (see later on).

Eq.(1) is an approximation of the more general equation expressing the second moment. In fact, one can more properly write:

$$\text{FWHM} \propto \sqrt{(<\Delta\nu^2>_d + <\Delta\nu^2>_m)} \qquad (2)$$

where $<\Delta\nu^2>_d$ is the intrinsic second moment due to dipolar interactions among nuclei, and $<\Delta\nu^2>_m$ is the second moment of the local frequency-shift distribution (due to nearby electronic moments) at the different proton sites of all molecules. To establish the correlation among the SQUID susceptibility $\chi_{SQUID}$ and the linewidth data, from the NMR spectra one can deduce the local susceptibility from the second moment $<\Delta\nu^2>_m$:

$$\chi_{loc} \propto \sqrt{<\Delta\nu^2>_m} \,/\, H \propto \sqrt{(FWHM)^2 - <\Delta\nu^2>_d} \,/\, H \qquad (3)$$

As further step, the existence of a proportionality of $\chi_{loc}$ to the macroscopic susceptibility $\chi_{SQUID}$ (measured with a SQUID magnetometer) can be verified with rescaling $\chi_{loc}$ by a factor $A$, so that: $\chi_{loc} = A\sqrt{<\Delta\nu^2>_m}/H$. As can be seen from Figs 10 and 11, $\chi_{loc}$ rescaled by $A$ and $\chi_{SQUID}$ coincide within the experimental errors in the range $1.5 < T < 30$ K.

It should be remarked that the local $\chi_{loc}$ can be calculated in principle taking into account the dipolar coupling constants averaged over all protons and all orientations. In Eq.(3), for $<\Delta\nu>_d$ we have used the values 37 kHz for [Cu$_6$Fe] and 40 kHz for [Cu$_6$Co], estimated by extrapolating the experimental FWHM vs H curve, collected at $T = 77$ K, to $H = 0$ (see Figs. 8 and 9). We chose the data at $T = 77$ K as the experimental linewidth at relatively high temperatures ($T > 150$K) can be narrowed by molecular motions.

We have fitted both local susceptibility and SQUID susceptibility with Curie-Weiss law $\chi T = C \cdot T/(T - \theta)$, using $C = 2.71$ cm$^3$mol$^{-1}$K, $\theta = 0.07$ K (accounting for a weak ferromagnetic interaction between Cu) for Cu$_6$Fe, and $C = 2.44$ cm$^3$mol$^{-1}$K, $\theta = -0.56$ K (accounting for a weak antiferromagnetic interaction between Cu) in the case of Cu$_6$Co. Hence, both bulk and local magnetic susceptibilities suggest that a weak ferromagnetic interaction between $Cu^{II}$, via $Fe^{III}$, in [$Cu_6Fe$], and a weak antiferromagnetic interaction between $Cu^{II}$, through $Co^{III}$, in [$Cu_6Co$] occur.

# Experimental details

**Syntheses**. [$Cu_2(saldmen)_2(\mu$-$H_2O)(H_2O)_2$]($ClO_4$)$_2$·$2H_2O$ **1**. To a methanolic solution (10 mL) containing 0.15 mmol of salicylaldehyde and 0.15 mmol of N,N-dimethylethylenediamine, 5 mL aqueous solution containing 0.15 mmol Cu($ClO_4$)$_2$·$6H_2O$ were added, under stirring. Violet-blue crystals of **1** formed within several days, which have been isolated by filtration. Selected IR bands (KBr, cm$^{-1}$): 3430m, 2924m, 1637s, 1600m, 1449wm, 1117vs, 1086vs, 1052vs, 627m.

[{$Cu^{II}$(saldmen)($H_2O$)}$_6${$Fe^{III}$(CN)$_6$}](ClO$_4$)$_3$·$8H_2O$ **2**. To a acetonitrile-water (1:1) solution (20 mL) containing 0.3 mmol of [$Cu_2(saldmen)_2(\mu$-$H_2O)(H_2O)_2$](ClO$_4$)$_2$·$2H_2O$, 10 mL acetonitrile-water (1:1) solution containing 0.1 mmol of K$_4$[Fe(CN)$_6$] were added under stirring. Green crystals suitable for X-ray diffraction were obtained directly from the reaction mixture, by slow evaporation of the filtrate at room temperature. Selected IR bands (KBr, cm$^{-1}$): 3560m, 3400m, 2925wm, 2104s, 1628s, 1590m, 1444wm, 1400wm, 751wm.

[{$Cu^{II}$(saldmen)($H_2O$)}$_6${$M^{III}$(CN)$_6$}](ClO$_4$)$_3$·$8H_2O$ (M = Co **3**, M = Cr **4**). To a acetonitrile-water (1:1) solution (20 mL) containing 0.3 mmol of [$Cu_2(saldmen)_2(\mu$-$H_2O)(H_2O)_2$](ClO$_4$)$_2$·$2H_2O$, 10 mL acetonitrile-water (1:1) solution containing 0.1 mmol of K$_3$[Co(CN)$_6$] (or K$_3$[Cr(CN)$_6$]) were added under stirring. Green crystals of **3** and **4**, suitable for X-ray diffraction, were obtained directly from the reaction mixtures, by slow evaporation at room temperature. Selected IR bands (KBr, cm$^{-1}$), **3**: 3399m, 2919wm, 2184s, 1638s, 1601m, 1450wm, 1410wm, 1094m, 761wm, 621wm; **4**: 3416m, 2928wm, 2183s, 1638s, 1601m, 1449wm, 1314wm, 1094m, 762wm, 621wm.

**X-Ray Crystallography**. Details about data collection and solution refinement are given in Table 4. X-ray diffraction measurements were performed on a STOE II Imaging Plate System both operating with a Mo-Kα ($\lambda = 0.71073$ Å) X-ray tube with a graphite monochromator. The structures were solved (SHELXS-97) by direct methods and refined (SHELXL-97) by full-matrix least-square procedures on $F^2$. All non-H atoms were refined anisotropically. Crystal data and details of the refinement for the two compounds are listed inTable 3.

**Magnetic measurements** were performed using a Cryogenics Squid S600 magnetometer with applied field of 0.1 T. To avoid possible orientation effects, microcrystalline powders were pressed in pellets. The data were corrected for sample holder contribution and diamagnetism of the sample using Pascal constants.

**NMR Spectra**. $^1$H NMR spectra measurements on polycrystalline $Cu_6Fe$ and $Cu_6Co$ were carried out with a standard TecMag Fourier transform pulse NMR spectrometer using short π/2-π/2 radio frequency (r.f) pulses (length~2 μs) in the temperature range 1.6 K to 300 K at two applied magnetic fields, $H = 0.5$ T and 1.5 T.[20] The high temperature NMR spectra, where the entire line could be irradiated with one r.f pulse, have been obtained from the Fourier transform of the half echo spin signal. The low temperature (1.6 - 20K) spectra have been built by convoluting the lines resulting from several Fourier transforms of each spectrum collected at different values of the irradiation frequency, keeping the external magnetic field constant.

**Acknowlegdements**.

This work was financially supported by the CNCSIS (PNII – IDEI-1912/2009, and by NE MAGMANet, NMP3-CT-2005-515767.

**Caption to the figures**

**Fig. 1** Perspective view of the binuclear complex **1**, along with the atom numbering scheme.

**Fig. 2** View of the heptanuclear cation in crystal **2**, along with the atom numbering scheme.

**Fig. 3** View of the octameric water cluster in crystal **2** (a); Packing diagram for crystal **2**, showing the water clusters sandwiched between the heptanuclear cations.

**Fig. 4** (a) Experimental $\chi T$ vs. T curve for **3**. In the inset the corresponding Curie-Weiss plot along with best ft curve. (b) Experimental $\chi T$ vs. T curve for **2** and best fit obtained with parameters reported in the text. (c) Experimental $\chi T$ vs. $T$ curve for **4** and best fit obtained with parameters reported in the text.

**Fig. 5.** $^1$H NMR spectra at various temperatures for **2** at $H = 1.5$ T. The red line is a simulation of the spectra through a sum of two Gaussian lines.

**Fig. 6**. $^1$H NMR spectra at various temperatures for **3** at $H = 1.5$ T. The red line is a simulation of the spectra through a sum of two Gaussian lines.

**Fig. 7**. Temperature dependence of FWHM (obtained by considering the central line only) in **2** at $H = 1.5$ T, and in **3** at $H = 1.5$ T and 0.5 T.

**Fig. 8**. Field dependence of FWHM at $T = 300$ K, 77 K and 4.2 K in **2**. The solid lines are linear fit to the data (see text).

**Fig. 9**. Field dependence of FWHM at $T = 300$ K, 77 K in **3**. The solid lines are linear fit to the data (see text).

**Fig. 10.** Temperature dependence of $\chi_{loc}T$ (renormalized) and $\chi_{SQUID}T$ in **2**, fitted by means of Curie-Weiss law (solid line), with fitting parameters given in the text.

**Fig. 11**. Temperature dependence of $\chi_{loc}T$ (renormalized) and $\chi_{squid}T$ in **3**, fitted with Curie-Weiss law (solid line) with fitting parameters given in the text.

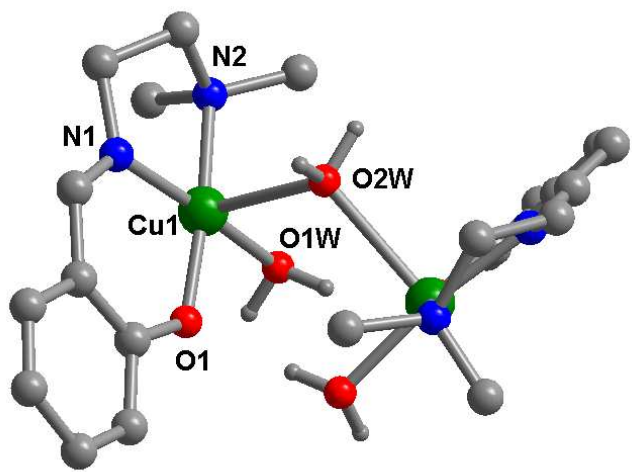

**Figure 1**

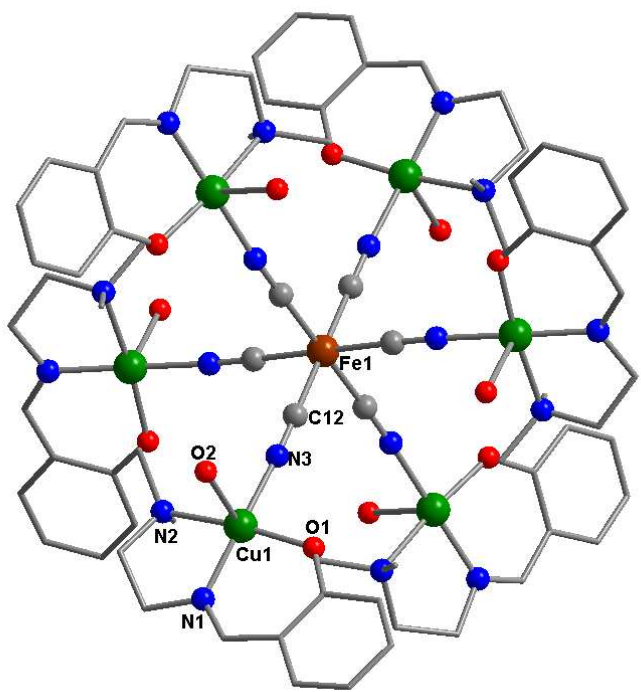

**Figure 2**

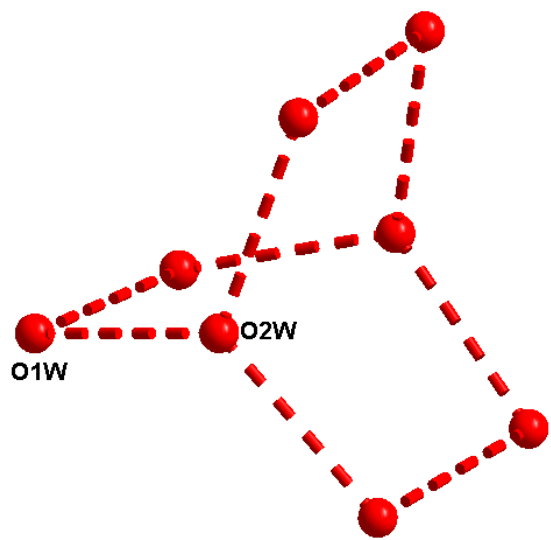

(a)

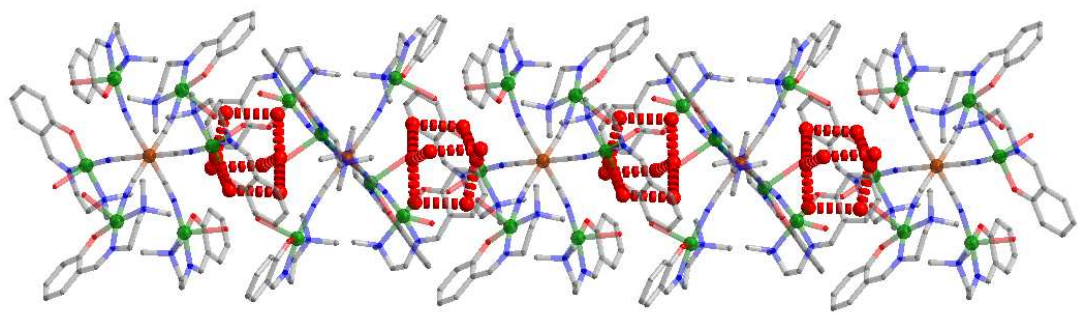

(b)

**Figure 3**

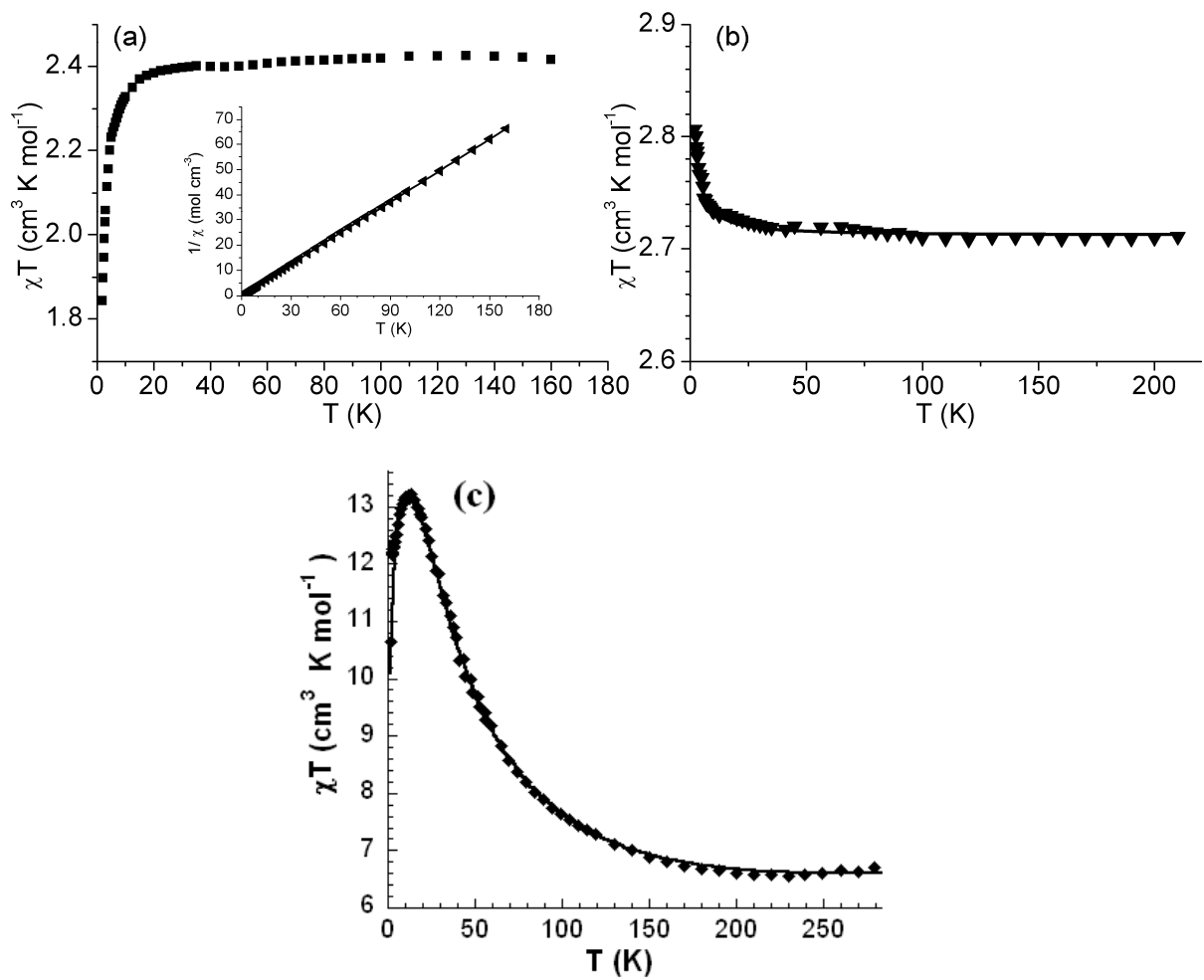

**Figure 4**

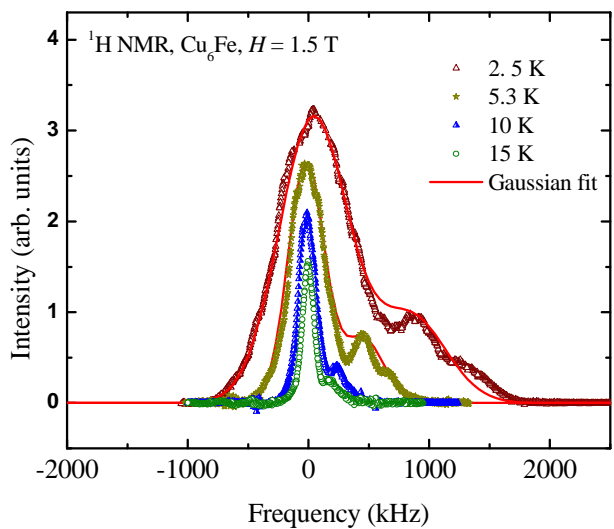

**Figure 5**

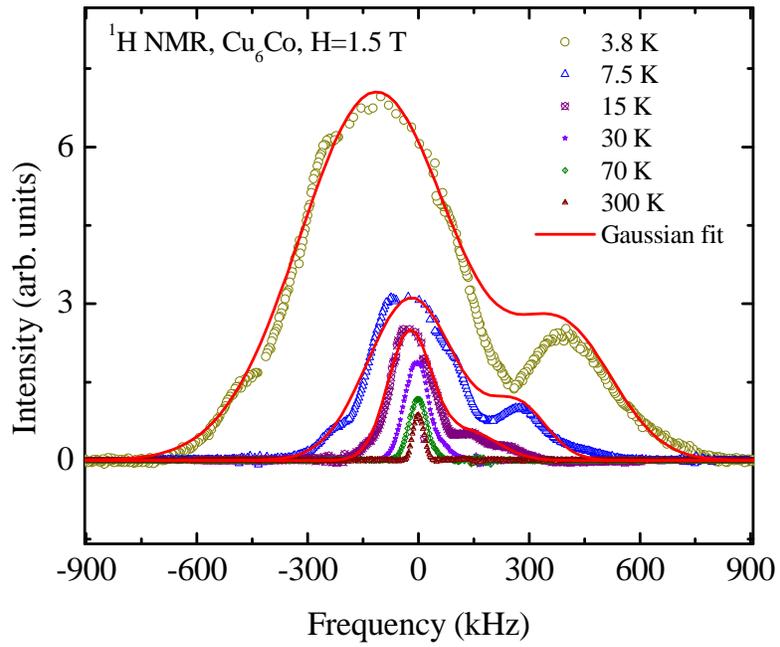

**Figure 6.**

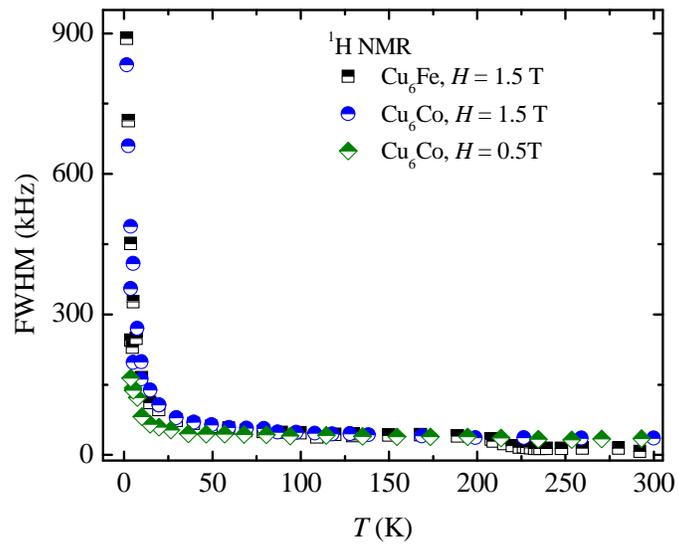

**Figure 7.**

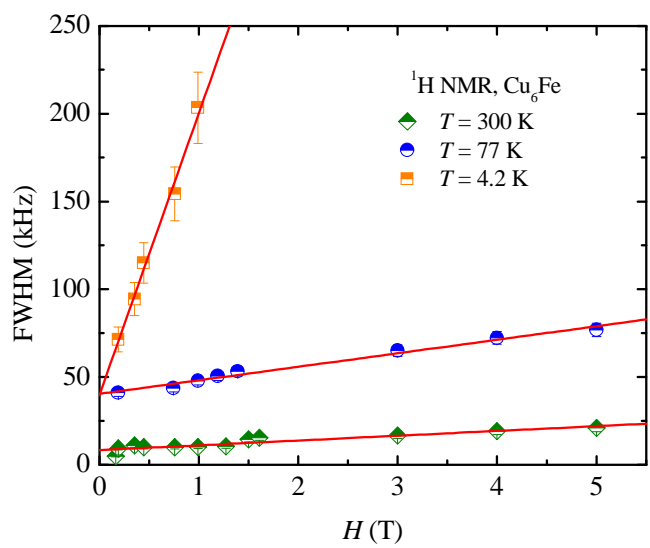

**Figure 8.**

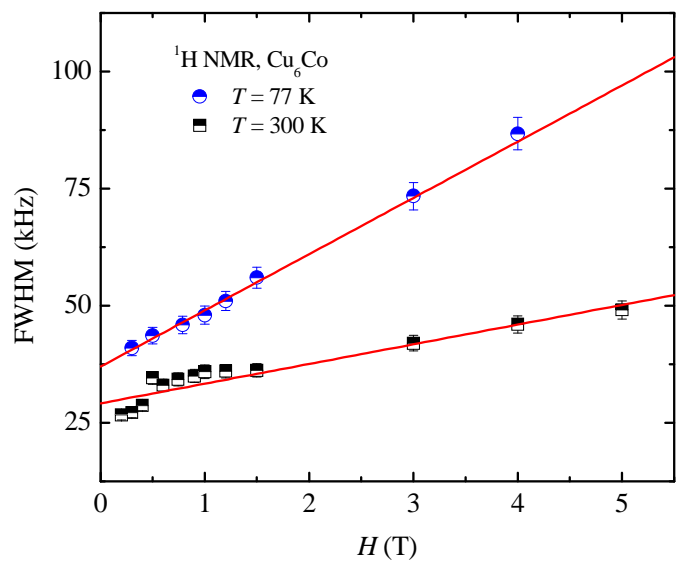

**Figure 9.**

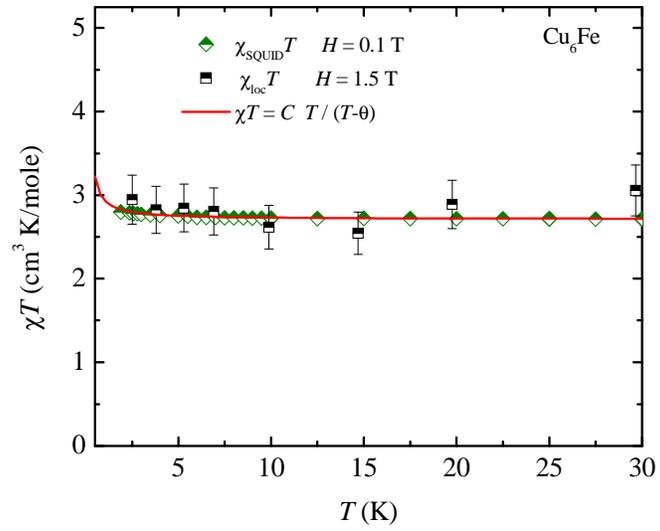

**Figure 10.**

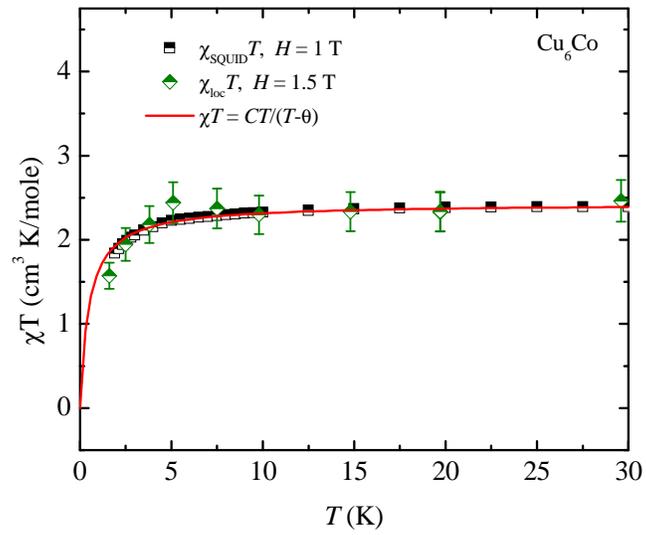

**Figure 11.**

**Table 1.** Selected bond distances (Å) and angles (°) for compounds **1**, **2**, **3** and **4**.

| | **1** | | **2** | | **3** | | **4** | |
|---|---|---|---|---|---|---|---|---|
| Cu1 O1 | 1.904(3) | C12 N3 | 1.154(7) | C12 N3 | 1.149(8) | C12 N3 | 1.148(6) | |
| Cu1 N1 | 1.937(4) | C12 Fe1 | 1.900(6) | C12 Co1 | 1.888(7) | C12 Cr1 | 2.045(5) | |
| Cu1 O1W | 1.966(4) | Cu1 O1 | 1.913(4) | Cu1 O1 | 1.927(5) | Cu O1 | 1.926(4) | |
| Cu1 N2 | 2.037(4) | Cu1 N1 | 1.923(5) | Cu1 N1 | 1.940(6) | Cu N1 | 1.940(4) | |
| Cu1 O2W | 2.477(3) | Cu1 N3 | 1.932(5) | Cu1 N3 | 1.973(6) | Cu N3 | 1.964(4) | |
| O1 Cu1 N1 | 93.34(16) | Cu1 N2 | 2.091(5) | Cu1 N2 | 2.085(6) | Cu N2 | 2.089(4) | |
| O1 Cu1 O1W | 89.16(14) | Cu1 O2 | 2.395(6) | Cu1 O2 | 2.358(6) | Cu O2 | 2.380(4) | |
| N1 Cu1 O1W | 171.28(16) | O1 Cu1 N1 | 92.58(19) | O1 Cu1 N1 | 92.4(2) | O1 Cu N1 | 92.44(17) | |
| O1 Cu1 N2 | 174.29(16) | O1 Cu1 N3 | 90.40(18) | O1 Cu1 N3 | 90.6(2) | O1 Cu N3 | 90.43(16) | |
| N1 Cu1 N2 | 85.0(2) | N1 Cu1 N3 | 172.7(2) | N1 Cu1 N3 | 171.0(3) | N1 Cu N3 | 170.84(18) | |
| O1W Cu1 N2 | 91.72(18) | O1 Cu1 N2 | 172.7(2) | O1 Cu1 N2 | 169.8(2) | O1 Cu N2 | 170.13(19) | |
| O1 Cu1 O2W | 90.08(12) | N1 Cu1 N2 | 83.9(2) | N1 Cu1 N2 | 83.9(2) | N1 Cu N2 | 84.41(18) | |
| N1 Cu1 O2W | 94.27(14) | N3 Cu1 N2 | 92.4(2) | N3 Cu1 N2 | 91.7(2) | N3 Cu N2 | 91.31(17) | |
| O1W Cu1 O2W | 94.08(13) | O1 Cu1 O2 | 93.9(3) | O1 Cu1 O2 | 97.7(2) | O1 Cu O2 | 97.51(18) | |
| N2 Cu1 O2W | 95.48(14) | N1 Cu1 O2 | 95.1(2) | N1 Cu1 O2 | 95.5(2) | N1 Cu O2 | 96.73(18) | |
| | | N3 Cu1 O2 | 91.3(2) | N3 Cu1 O2 | 92.4(2) | N3 Cu O2 | 91.51(16) | |
| | | N2 Cu1 O2 | 92.8(3) | N2 Cu1 O2 | 92.1(2) | N2 Cu O2 | 92.2(2) | |

**Table 2**. Geometrical parameters of the hydrogen bonds in compound **2**

| | |
|---|---|
| O2···O1w | 2.83 |
| O1w···O2w | 2.43 |
| O1w···O1w[i] | 2.77 |

(i) –$x$, -$x$+$y$, 0.5-$y$

**Table 3**. Crystallographic data, details of data collection and structure refinement parameters for compounds **1**, **2**, **3** and **4**.

| Compound | **1** | **2** | **3** | **4** |
| --- | --- | --- | --- | --- |
| Chemical formula | $C_{22}H_{40}Cl_2Cu_2N_4O_{15}$ | $C_{72}H_{90}Cl_3Cu_6FeN_{18}O_{32}$ | $C_{72}H_{90}Cl_3CoCu_6N_{18}O_{32}$ | $C_{72}H_{90}Cl_3CrCu_6N_{18}O_{32}$ |
| $M$ (g mol$^{-1}$) | 798.56 | 2263.06 | 2266.14 | 2259.21 |
| Temperature, (K) | 150(2) | 293 | 293 | 293 |
| Wavelength, (Å) | 0.71073 | 0.71073 | 0.71073 | 0.71073 |
| Crystal system | *orthorhombic* | *hexagonal* | *hexagonal* | *hexagonal* |
| Space group | *Fdd2* | *R-3c* | *R-3c* | *R-3c* |
| $a$(Å) | 17.5460(11) | 27.8777(16) | 27.9545(19) | 28.100 |
| $b$(Å) | 37.8030(11) | 27.8777(16) | 27.9545(19) | 28.100 |
| $c$(Å) | 9.994(2) | 21.3690(13) | 21.3938(16) | 21.656 |
| $\alpha$(°) | 90 | 90 | 90 | 90 |
| $\beta$(°) | 90 | 90 | 90 | 90 |
| $\gamma$(°) | 90 | 120 | 120 | 120 |
| $V$(Å$^3$) | 6628.9(15) | 14382.3(15) | 14478.4(18) | 14808.9 |
| $Z$ | 8 | 6 | 6 | 6 |
| $D_c$ (g cm$^{-3}$) | 1.600 | 1.568 | 1.559 | 1.520 |
| $\mu$ (mm$^{-1}$) | 1.516 | 1.622 | 1.633 | 1.538 |
| $F(000)$ | 3296 | 6930 | 6936 | 6918 |
| Goodness-of-fit on $F^2$ | 1.002 | 0.982 | 1.078 | 0.958 |
| Final $R1$, $wR_2$ [$I>2\sigma(I)$] | 0.0440, 0.1002 | 0.0596, 0.1546 | 0.0682, 0.1349 | 0.0596, 0.1081 |
| $R1$, $wR_2$ (all data) | 0.0643, 0.1055 | 0.1056, 0.1828 | 0.1321, 0.1581 | 0.1553, 0.1359 |
| Largest diff. peak and hole (eÅ$^{-3}$) | 0.558, -0.614 | 0.526, -0.756 | 0.703, -0.682 | 0.572, -0.545 |